\documentclass[final,3p,times,twocolumn,numbers]{elsarticle}

\usepackage{amssymb}

\usepackage{epsfig}
\usepackage{graphicx}
\usepackage{bm}
\usepackage{color}
\usepackage{amsmath,amsfonts} 
\usepackage{tikz}
\usepackage{subfigure}
\usepackage{multirow}
\usepackage{booktabs}
\usepackage{wrapfig}
\usepackage{verbatim}
\usepackage{hyperref}

\newcommand{\be}{\begin{eqnarray}}
\newcommand{\ee}{\end{eqnarray}}

\journal{Physics Letters B}

\begin{document}

\topmargin -1cm
\newlength{\axnucl}

\begin{frontmatter}



\title{Many-body effects in (p,pN) reactions within a unified approach}



\author[IST,C2TN]{R.~Crespo\corref{correspondingauthor}}
\cortext[correspondingauthor]{Corresponding author}
\ead{raquel.crespo@tecnico.ulisboa.pt}

\author[DFFCUL]{A.~Arriaga}
\author[ANL]{R.B.~Wiringa}
\author[CFTC,DFFCUL]{E.~Cravo}
\author[IST,C2TN]{A.~Mecca}
\author[ITPA]{A.~Deltuva}

\address[IST]{Departamento de F\'{\i}sica, Instituto Superior T\'ecnico, Universidade de Lisboa, Av.~Rovisco~Pais~1, 1049-001, Lisboa, Portugal}

\address[C2TN]{Centro de Ci\^encias e Tecnologias  Nucleares, Universidade de Lisboa,
Estrada Nacional 10, 2695-066 Bobadela, Portugal}

\address[DFFCUL]{Departamento de F\'{\i}sica, Faculdade de Ci\^encias, Universidade de Lisboa, Campo Grande, 1749-016~Lisboa, Portugal}

\address[ANL]{Physics Division, Argonne National Laboratory, Argonne, Illinois 60439, USA}

\address[CFTC]{Centro de F\'{\i}sica Te\'{o}rica e Computacional, Faculdade de Ci\^encias, Universidade de Lisboa, Campo Grande, 1749-016~Lisboa, Portugal}

\address[ITPA]{Institute of Theoretical Physics and Astronomy, Vilnius University, 
Saul\.etekio al. 3, LT-10257  Vilnius, Lithuania}

\begin{abstract}
We   study knockout reactions with proton probes within a theoretical framework where {\it ab initio} Quantum Monte Carlo wave functions are combined with the 
 Faddeev/Alt-Grassberger-Sandhas  few-body reaction formalism. New Quantum Monte Carlo wave functions are used to describe $^{12}$C, yielding, for the first time, results consistent with the experimental point rms radii, electron scattering data and (p,2p) total cross sections data. 
Our results for $\mathrm{A}\leq 12$ and $(N-Z) \leq 3$ nuclei show that  the theoretical ratios between the (i) 
 {\it ab initio} and Mean Field Approximation theoretical  cross sections, $\mathcal{R}_\sigma$,  (ii)   corresponding ratios between the spectroscopic factors,  $\mathcal{R}_\Sigma$, 
summed over states below particle emission, depend moderately on the nucleon separation energy S$_{\rm N}$.
These ratios  are determined by a delicate interplay between the radii of the parent and the residual nuclei and the nucleon separation energy, and
were found to be always smaller for the knockout of the more correlated deficient species nucleon.  In the case 
of  the symmetric $^{12}$C nucleus, the theoretical ratios  still appear to indicate that protons are more correlated than neutrons.

\end{abstract}




\end{frontmatter}


\section{Introduction}


The mean field approach (MFA) to particle systems  has played an important role in atomic physics for describing the periodic table of elements and in nuclear physics for explaining many properties of nuclei, such as the origin of the magic numbers leading to additional stability.

Nevertheless, one of the goals of Nuclear Physics is to describe simultaneously, and along the nuclear landscape, nuclear binding, structure, electromagnetic and weak transitions, 
as well as reactions with electroweak and nuclear probes  based on a microscopic description of the interaction between individual nucleons.

A formidable theoretical effort has been performed in developing many body  and cluster approaches 
to describe nuclei and their application in the study of reactions \citep{wiringa14,Carlson15,Dickoff05,Navratil08,Natasha13, Barranco17,Jensen} .
Strong deviations between these models and the MFA  indicate the presence of non-trivial many-body effects, being 
interpreted as due to nuclear correlations. 
Many body  {\it ab initio} calculations of nuclear structure have demonstrated the need to go beyond the simplified MFA  and to
consider models with explicit nucleon-nucleon (NN) and three-nucleon (NNN) interactions  and NN and NNN correlations \citep{wiringa14,Carlson15}, 
in particular neutron-proton correlations entirely absent in the MFA \citep{Phandharipande,nature}. 

In parallel, for more than 30 years an extensive experimental program,  in particular nucleon knockout reactions with electron and nuclear probes,
has been devoted to the study of the failure of the MFA
\citep{lapikas93, lapikas99, Rhoe,Subedi,Kramer01, Tostevin-02, Tostevin-14, grinyer12, leyla, paloma,Nollett11, Lee, flavignyPRL}.

In  most calculations, the extracted information depends more or less strongly on the uncertainties of  the reaction formalism and  its associated interactions.
Moreover, the variety of models,  methods and energy regimes makes difficult to extract a consistent explanation of the MFA inadequacy to describe nuclear structure and
to value the importance and nature of the correlations.

The interpretation of the single nucleon knockout from an A-body nucleus has been relying on  the one-nucleon spectroscopic overlap. 
This is defined as the inner product of the A parent nucleus wave function (WF)
 and the fully antisymmetrized $\mathrm{A}\!-\!1$ residual nucleus (core, C) plus the knockout nucleon WF.  
For a given state of the residual nucleus, this overlap is a superposition of different nucleon angular momentum channels, $\ell j$, satisfying the appropriate triangular relations \citep{brida11}.
The spectroscopic factor (SF) for a given transition is given by the integral of the overlap function in each angular momentum channel.

The analysis of earlier (e,e'p) knockout experiments has been used to provide information on the one-nu\-cle\-on spectroscopic overlaps at low momentum and for low-lying energy states of the residual nucleus. The experimentally extracted SFs were found to be reduced  with respect to the MFA ones \citep{lapikas93,lapikas99}. 

The nucleon knockout for composite projectiles and target nuclei (called one-nucleon removal in the literature)  
has been also analyzed extensively \citep{Tostevin-14,grinyer12} and refs therein.  
The ratio between the inclusive experimental and  the MFA theoretical  cross sections,  R$_S$, has been found to be smaller than unity and to have a
strong dependence on the asymmetry parameter $\Delta S$, a measure of the asymmetry of the neutron and proton binding. This has been interpreted as additional correlations in strongly asymmetric (N-Z) systems.
 
 Concurrently, (p,pN)  reactions on Oxygen, Carbon and Nitrogen isotopes with $ -2 \leq (N-Z) \leq 14$ \citep{leyla,paloma,Moro}, 
 and  transfer studies of (d,t) and (d,$^3$He) on $^{14,16,18}$O   \citep{flavignyPRL}  and of  (p,d)  on  $^{34,46}$Ar \citep{Lee} 
 have revealed small and nearly constant   $\mathcal{R}_S$  as a function of $\Delta S$. 

Theoretical calculations of one-nucleon spectroscopic overlaps for asymmetric parent nuclei, $^{14,16,22,24,28}$O showed that SFs calculated with a microscopic coupled cluster model
are quenched relatively to the MFA ones, the quenching being particularly important for the knockout of deficient species nucleons in strongly asymmetric nuclei
\citep{ Jensen}.   On the other hand, a small dependence of this quenching on the nucleon binding was found \citep{Dickoff05}.

Conflicting results did follow from this vast theoretical and experimental work. 
A consistent analysis of available experimental data,  for all open reaction channels as well as different probes, with state-of-the art theory is lacking and of utmost importance for the understanding of nuclear structure along the nuclear landscape. Furthermore, it is essential to meet the challenges of new experimental developments and multiphysics  research \citep{Bortignon-16}.

%

In this letter our goal is to contribute to a unified theoretical approach built on {\it ab initio} WFs, which can be used as a common input to transfer and nucleon knockout reactions with electron and proton probes.
We aim  to shed light  on the failure of the MFA to describe this type of reactions and to provide an understanding of (i) the ratios  $\mathcal{R}_S$, (ii)  the ratios between 
the  {\it ab initio} and MFA theoretical  cross sections, $\mathcal{R}_\sigma$, (iii) their relation with the corresponding ratios between the spectroscopic factors,  $\mathcal{R}_\Sigma$, and (iv) their 
behaviour as a function of the separation energy of the knocked out nucleon.
Our analysis of (p,pN) reactions with light nuclei will contribute to the construction of a unified interpretation of nucleon knockout reactions along the nuclear landscape, including the (p,pN) experimental data collected at the R3B-LAND setup at GSI \citep{leyla,paloma,panin}.

To achieve the above  goal, we use Quantum Mon\-te Carlo (QMC) methods to solve the many body Schr\"o\-ding\-er equation,  which are state-of-the-art techniques used in various subfields of physics, such as molecular, atomic and nuclear physics. 
In particular, in the latter, {\it ab initio} QMC calculations \citep{wiringa14} have been used to interpret sucessfully transfer reactions \citep{Nollett11} and (e,e'p) experimental data \citep{lapikas99}. We also aim to test, for the first time, the ability of QMC calculations in describing (p,pN) reactions.

\section{Formalism}
%
%

The main assumption of  our model is that the knockout/breakup operator does not act on the internal structure of the residual nucleus,
 making the spectroscopic overlap between the parent and residual nuclei the key nuclear structure input. These overlaps are calculated from QMC many-body wave functions and then incorporated in the state-of-the-art Faddeev/Alt-Grassberger-Sandhas (F/AGS) to solve the resulting three-body scattering problem \citep{theorFad,alt:67a}.
Our assumption is supported by recent work \cite {Deltuva-19}, where it has been shown that
dynamical core excitation effects in (p,pN) reactions are small, to a good approximation validating the factorization of the cross section into the single-particle cross section, defined below, and the corresponding SF. This enables direct spectroscopic information from the comparison between  experimental and theoretical cross sections.

The F/AGS allows a consistent and simultaneous  treatment of all open channels, providing an exact solution of the
three body scattering problem for an assumed three-body Hamiltonian. 
This formalism includes all multiple scattering terms, contrary to other scattering frameworks that rely on assumed exact cancellations between multiple scattering terms \citep{Crespo-19}.
It has been used recently in several exploratory studies of (p,pN) reactions
\citep{paloma,Crespo-19,cravo,crespoJPG} and it is able to model the experimental transverse momentum distributions \citep{Crespo-19}.

We use the F/AGS  in a nonrelativistic form since consistent treatment of relativistic kinematics and dynamics 
in Refs.~\citep{Witala06,Witala11} indicates only a small relativistic effect for the total three-body breakup cross section,  less than 10 $\%$ in our energy regime of interest.

The reaction formalism requires  three pair interactions. We take the realistic NN AV18 for the proton-nucleon pair.
For the the N-C and p-C pair interactions
we consider the Koning-Delaroche (KD) optical parametrization \citep{KD} used in preliminary calculations 
\cite {crespoJPG} and the Cooper \cite {cooper} for $^{12}$C, a global parametrization developed for medium-heavy nuclei and in particular for A=12 that reproduce the elastic scattering data.
From comparison with  other parametrizations provided in \citep{Crespo-19}  we estimate the uncertainty on the cross sections associated with optical parametrizations of 15 $\%$.

In our approach, the spectroscopic overlaps are calculated from the QMC many-body WFs generated 
using the NN Argonne V18  and the NNN Urbana X (AV18+UX) potentials  \citep{Carlson15}. 
We consider variational Monte Carlo (VMC) overlaps for p- and n-knockout from  $^{9}$Li, $^{10}$Be and $^{12}$C nuclei. 
For the nuclear structure input we use new improved VMC WFs for $^{12}$C with [444] and [4431] spatial symmetries
($^{11}$B with [443] and [4421]) as specified in Young diagram notation~\citep{wiringa06}.
Preliminary results with the Norfolk local chiral potential 
NV2+3-Ia*~ \citep{piarulli18,baroni18} show minor variations in the SFs of 5 $\%$ with respect to AV18+UX.
The GFMC SFs for the $^7$Li parent  and residual $^6$Li  overlaps 
agree fairly well with the VMC ones, supporting the use of the VMC overlaps for the study of these reactions.
We take the VMC and Green's function Monte Carlo (GFMC) overlaps for the $^{7}$Li  parent nucleus from    Ref.~\citep{brida11},   which are able to describe the (e,e'p)  reaction \citep{lapikas99}. 

We have performed a convenient parametrization of the QMC overlaps using the procedure described in Ref.~\citep{brida11}, which incorporates the adequate asymptotic behavior. 
With a correct tail by construction, these overlaps can be validated indirectly by comparing 
the experimental values of the point proton rms radii ($r_p$) with the ones obtained from the QMC WFs from which the overlaps are calculated.
 In Table \ref{Table-Radii}, we present the VMC and the experimental values for $r_p$ \citep{Vries87,Lu13}   for all the studied nuclei, which exhibit
 a quite good agreement.

We also consider an MFA where only the (A-1)+N configuration is present in the parent nucleus state space. 
The overlaps are then obtained as solutions of the one-body Schr\"odinger equation with an effective average interaction.
We use  a Woods-Saxon potential with standard radius and depth adjusted to the separation energy of the removed nucleon and no antisymmetrization is considered.
As discussed below, from the overlaps, the SFs are the crucial quantities in the cross section, and for this reason we take the SFs calculated from the more sophisticated  effective interaction
of Cohen-Kurath (CK) with the  well known center of mass (c.m.) correction,  given by $A/(A-1)$, Ref.~\citep{CK,cm_corr}.


%
%
The theoretical SFs for each  structure  model, ${\mathcal M}$, (QMC and MFA) are denoted here as $Z ^{\rm i}({\mathcal M})$, where $i$ identifies the energy and the angular momentum of the residual nucleus, as well as the nucleon angular momentum channels, with the sum ${\Sigma({\mathcal M}) = \Sigma_i Z^{\rm i}({\mathcal M})}$.

The theoretical inclusive cross section $\sigma_{\rm th}({\mathcal M})$
is obtained as the weighted sum ${\sigma_{\rm th}({\mathcal M}) = \sum_i Z ^{\rm i}({\mathcal M}) \sigma_{\rm sp}^{\rm i}({\mathcal M})}$
where  the single-particle cross sections $\sigma_{\rm sp}^{\rm i}({\mathcal M})$ are computed using the overlaps normalized to unity. 

%
%

%
%
\begin{figure}[htb]
\includegraphics[angle=-90,width=0.47\textwidth]{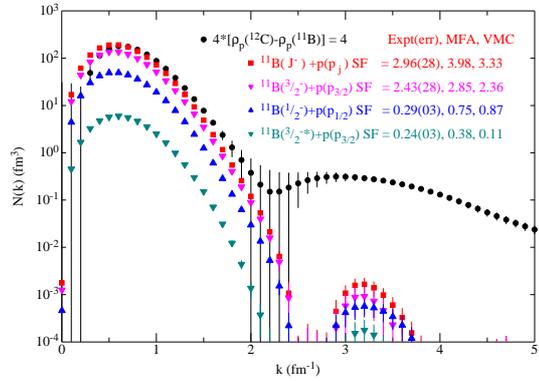}
\caption{\label{12Coverlaps}VMC overlaps in momentum space  $N(k)$ calculated  for the low-lying  states of $^{11}$B.
Also shown in black is the difference of $^{12}\mathrm{C}$ and $^{11}\mathrm{B}$  proton momentum distributions multiplied by 4 (the total number of protons in a p-shell in the Independent Particle Model).}
\end{figure}
%
%

\section{Results}

We start by evaluating the one-nucleon spectroscopic overlaps for the parent nucleus $^{12}$C, for which  there are (p,2p) data \citep{panin}.
The overlaps in momentum space 
are represented in Fig.~\ref{12Coverlaps}, along with
 the difference between the VMC $^{12}\mathrm{C}$ and  $^{11}\mathrm{B}$ proton momentum distributions. This difference exhibits a significant high-momentum tail, where about 15$\%$  of the protons have momenta above 1.4 fm$^{-1}$, unaccountable in any MFA.
 The result for $^{12}\mathrm{C}$, shown here for the first time, is consistent with high momentum electron scattering analysis  \citep{Rhoe}, supporting the VMC WF from which both the momentum distributions and the spectroscopic overlaps are generated and, therefore, corroborating our nuclear structure model. 
 Interesting to say that the dominant source of this high momentum tail is the NN tensor force, coming from the one-pion-exchange potential, with a further significant contribution from the NNN force with its two-pion-exchange terms.
 The experimental SFs, $Z^i_{\rm exp}$,  deduced from $^{12}\mathrm{C}$ (p,2p) at 400 MeV/u, 
are calculated dividing the experimental cross section of  Ref.~\citep{panin} by the $\sigma_{\rm sp}^{\rm i}({\rm VMC})$. 
These  are shown in the inset of  Fig.~\ref{12Coverlaps} along with the theoretical VMC and MFA ones.
The theoretical sums  $\Sigma({\mathcal M}) = \sum_i    Z ^{\rm i} ({\mathcal M})     $ and the experimental one 
for all final states of the residual nucleus  are also shown.
It is fair to say that, when compared to the MFA ones, the VMC SFs lead to a significant improvement.
In addition, their sum agrees reasonably well
with the deduced experimental sum ${ \Sigma_{\rm exp} }$(p,2p) = $2.96$ 
and moderately with those extracted from electron scattering  ${ \Sigma_{\rm exp} }$(e,e'p) = $2.18(15)(1.00)$ \citep{Steenhoven} and transfer
${ \Sigma_{\rm exp} }$(d,$^3$He) = $2.1(1.00)$ \citep{Kramer01}.
We also note that $\Sigma({\rm MFA})$ is very close to the sum of particles in the shell (before c.m.\ correction), the well known sum rule.
We have obtained the total theoretical cross section $\sigma_\mathrm{th}(\mathrm{QMC}) = 21.66\,\mathrm{mb}$, close to the experimental 
value of $\sigma_\mathrm{exp} = 19.2(18)(12)\,\mathrm{mb}$ \citep{panin}, with the ratio of experimental and theoretical  values being $0.886(10)$.
This result shows, for the first time, that   {\it ab initio} VMC  WFs combined with the Faddeev/AGS reaction formalism  predict  cross sections for (p,2p) from $^{12}\mathrm{C}$ 
 that agree fairly well with the experimental data.  In other words, this result shows the ability of QMC WFS to describe (p,pN) reactions,  within a remnant uncertainy
 due to optical parametrizations and relativity.

%
%
%
\begin{table}[htb]
\caption{\label{Table-Radii}Radii, nucleon separation energies and SFs for the ground states  of the   parent  $^\mathrm{A}\mathrm{X}$ and residual nucleus $^\mathrm{A-1}\mathrm{Y}$.}
\settowidth{\axnucl}{$^{10}\mathrm{Be}$}
{\footnotesize
\centering
\begin{tabular}{@{}*{2}{c@{}}*{5}{c@{\hspace{0.6em}}}c@{}c@{}}
\toprule
\multirow{3}{*}{$^{\rm A}{\rm X}$} & 
\multirow{3}{*}{$^{\rm A-1}{\rm Y}$} &
\multirow{3}{*}{$J^\pi$} &
$S_N$ &
$r_p$ &
$r_n$ &
$r_m$ &
$r_p$ &
$Z^i_{\scriptscriptstyle\mathrm{VMC}}/Z^i_{\scriptscriptstyle\mathrm{MFA}}$ \\
&&& (MeV) & (fm) & (fm) & (fm) & (fm) & SF \\
&&& exp & VMC & VMC & VMC & exp & g.s. \\
\midrule
$^7\mathrm{Li}$ && $3/2^+$ && 2.26 & 2.41 & 2.35 & 2.31(5) & \\
& $^6\mathrm{Li}$ & $1^+$ & 7.25 & 2.46 & 2.46 & 2.46 & 2.45(4) & 0.81 \\
& $^6\mathrm{He}$ & $0^+$ & 9.97 & 1.94 & 2.82 & 2.53 & 1.92(1) & 0.56 \\
\midrule
$^9\mathrm{Li}$ && $3/2^-$ && 2.07 & 2.45 & 2.33 & 2.11(5) & \\
& $^8\mathrm{Li}$ & $2^+$ & 4.06 & 2.13 & 2.44 & 2.33 & 2.20(5) & 0.96 \\
& $^8\mathrm{He}$ & $0^+$ & 13.94 & 1.83 & 2.79 & 2.58 & 1.84(2) & 0.67 \\
\midrule
$^{10}\mathrm{Be}$ && $0^+$ && 2.31 & 2.51 & 2.43 & 2.22(2) & \\
& $^9\mathrm{Be}$ & $3/2^+$ & 6.81 & 2.36 & 2.46 & 2.42 & 2.36(1) & 0.83 \\
& $^9\mathrm{Li}$ & $3/2^+$ & 19.64 & 2.07 & 2.45 & 2.33 & 2.11(5) & 0.60 \\
\midrule
$^{12}\mathrm{C}$ && $0^+$ && 2.37 & 2.37 & 2.37 & 2.32(1) & \\
& $^{11}\mathrm{C}$ & $3/2^-$ & 18.72 & 2.41 & 2.35 & 2.38 & - & 0.77 \\
& $^{11}\mathrm{B}$ & $3/2^-$ & 15.96 & 2.35 & 2.41 & 2.38 & 2.28(13) & 0.77 \\
\bottomrule
\end{tabular}
}
\end{table}
%
%
The calculated QMC and  MFA  SFs,  together with their sums $\Sigma$  and the
ratios    $\mathcal{R}_\Sigma$  (including c.m.\ correction)  are shown in {Table~\ref{Table-SF}.
Also shown  are the results with the partial sum over the final states of the residual nucleus below its breakup threshold 
 (called here below particle threshold, BPT).

%
%
\begin{table}[htb]
\caption{\label{Table-SF}
Total and BPT sums of SFs, $\Sigma$, and ratios $\mathcal{R}_{\Sigma}$. The MFA$^{\star}$ includes c.m.\ correction factors.}
\settowidth{\axnucl}{$3.568(22)$}
{\footnotesize
\centering
\begin{tabular}{@{}*{7}{c@{\hspace{0.8em}}}c@{}}
\toprule
\multirow{2}{*}{\normalsize $^\mathrm{A}\mathrm{X}$} & 
\multirow{2}{*}{\normalsize $^\mathrm{A-1}\mathrm{Y}$} &
&&
\multicolumn{2}{c}{$\Sigma(\mathcal{M})$} &
&
$\mathcal{R}_\Sigma$ \\
&&&&
\parbox{\axnucl}{\centering MFA} & 
\parbox{\axnucl}{\centering VMC} &
&
VMC/MFA$^{\star}$ \\
\midrule
\multirow{5}{*}{\normalsize $^7\mathrm{Li}$} &
\multirow{2}{*}{\normalsize $^6\mathrm{Li}$} &
& BPT & $1.016$ & $0.874(3)$ && $0.737(2)$ \\
&&&& $1.999$ & $1.606(10)$ && $0.689(4)$  \\
&
\multirow{2}{*}{\normalsize $^6\mathrm{He}$} &
& BPT & $0.592$ & $0.389(1)$ && $0.563(1)$ \\
&&&& $0.997$ & $0.733(3)$ && $0.630(2)$  \\
\midrule
\multirow{5}{*}{\normalsize $^9\mathrm{Li}$} &
\multirow{2}{*}{\normalsize $^8\mathrm{Li}$} &
& BPT & $1.313$ & $1.428(4)$ && $0.967(3)$ \\
&&&& $3.859$ & $3.597(14)$ && $0.829(3)$  \\
&
\multirow{2}{*}{\normalsize $^8\mathrm{He}$} &
& BPT & $0.847$ & $0.635(2)$ && $0.666(2)$ \\
&&&& $1.000$ & $0.785(3)$ && $0.698(3)$ \\
\midrule
\multirow{5}{*}{\normalsize $^{10}\mathrm{Be}$} &
\multirow{2}{*}{\normalsize $^9\mathrm{Be}$} &
& BPT & $2.356$ & $2.174(2)$ && $0.830(1)$ \\
&&&& $3.990$ & $3.568(22)$ && $0.805(5)$ \\
&
\multirow{2}{*}{\normalsize $^9\mathrm{Li}$} &
& BPT & $1.990$ & $1.597(5)$ && $0.722(2)$ \\
&&&& $1.990$ & $1.676(7)$ && $0.758(3)$ \\
\midrule
\multirow{5}{*}{\normalsize $^{12}\mathrm{C}$} &
\multirow{2}{*}{\normalsize $^{11}\mathrm{C}$} &
& BPT & $3.603$ & $3.234(17)$ && $0.823(16)$ \\
&&&& $3.980$ & $3.326(17)$ && $0.766(16)$ \\
&
\multirow{2}{*}{\normalsize $^{11}\mathrm{B}$} &
& BPT & $3.980$ & $3.333(17)$ && $0.768(16)$ \\
&&&& $3.980$ & $3.387(17)$ && $0.780(16)$ \\
\bottomrule
\end{tabular}
}
\end{table}
%

%
%
The  ratios $\mathcal{R}_\Sigma$ 
range from 0.6 to 0.8 being  consistent with Ref.~\citep{Phandharipande}. 
This reduction is due to the fact that the MFA considers only the $(\mathrm{A}\!-\!1)+\mathrm{N}$ partition for the parent nucleus wave function, therefore setting to unity the probability of finding this configuration inside the nucleus.
In contrast, the QMC overlaps are calculated from fully microscopic WFs for parent and residual nuclei, both normalized to unity.
This means that many other partitions are present in the parent nucleus WF leading to a probability associated with $(\mathrm{A}\!-\!1)+\mathrm{N}$ configuration smaller than unity. Consequently, the MFA SFs are  necessarily larger than the QMC and the experimental ones. This conclusion is  independent of the interaction models.
The ratios calculated with partial sums over the final states
of the residual nuclei,   ${\mathcal R}_{\Sigma}$ BPT, shown in the upper panel of Fig.~\ref{fig:theoratios}, differ significantly from those calculated with total sums, ranging from 0.5 to 1.
These results follow naturally from the fact that the spectroscopic strength is distributed among the states differently in the VMC and MFA formalisms.
The ${\mathcal R}_{\Sigma}$ BPT exhibit a  moderate dependance  on S$_{\rm N}$, for the considered small asymmetry 
$(N-Z) \leq 3$.
Nevertheless, the ratio ${\mathcal R}_{\Sigma}$  is always smaller for the knockout of the more correlated deficient species nucleon, the proton in these cases, in accordance with
previous findings \cite{Jensen,Tostevin-14}.

On the other hand, we have found that the overlaps, and consequently the SFs, are determined by a delicate interplay between the radii of the parent (A) and the residual $(\mathrm{A}\!-\!1)$ nuclei and the separation energy of the knockout nucleon.
To further explore this interplay, we present in Table I, for the ground states of parent and residual nuclei, the separation energy S$_{\rm N}$, the point proton and neutron rms radii, the matter radii, given by 
$r_{\rm m}=\sqrt{(Z r_{\rm p}^2+N r_{\rm n}^2)/{\rm A}}$, and the corresponding ratios of VMC/MFA SFs for the ground state as well. It is evident that there is no clear dependence of these ratios on the separation energy, for instance they do not decrease necessarily with increasing S$_{\rm N}$, reflecting the fact that the SFs do not probe exclusively the tail of the overlaps. In fact, these ratios are also determined by the proximity between the matter radii of parent and daughter nuclei, which tends to enhance the overlap. 
Being this interplay dealt differently in VMC and MFA calculations, it is clear the subsequence dependance of ${\mathcal R}_{\Sigma}$ on it.

%
%
\begin{figure}[htb]
\includegraphics[width= 0.38 \textwidth]{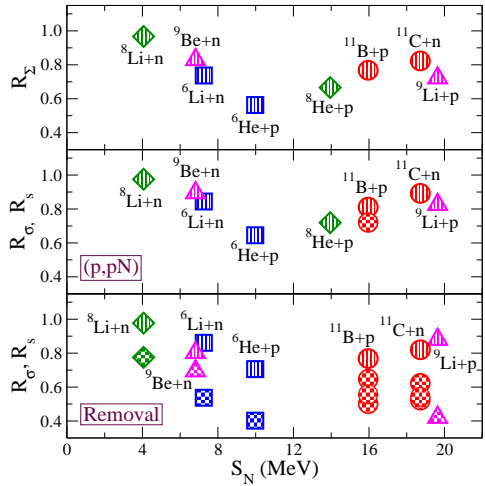}
\caption{ Theoretical ratios (QMC/MFA)  of sums  SFs (upper panel), (p,pN)  (middle pannel) and removal (lower panel)
cross sections  restricted to final states below particle threshold (BPT) of the residual nucleus and quenching factors, $\mathcal{R}_S$ (symbols filled with squares),
as a function of the nucleon separation energy.
The  theoretical single particle removal cross sections are taken from Refs.~\citep{Tostevin-02,grinyer12}. 
}
\label{fig:theoratios}
\end{figure}
%

%
%
We also calculate the  ratios between the QMC and  MFA  theoretical cross sections,
 but where the sum is restricted to states of the residual nucleus  BPT.
Since we are considering low-lying states of the residual nucleus and  we found only weak dependence of the $\sigma_{\rm sp}^{\rm i}$  on S$_{\rm N}$ \citep{cravo}, we expect  these ratios to be nearly independent of the choice of the global parametrization and  MFA details.
For the case of $^{12}$C we verified a model independence of these ratios since similar results are obtained with different parameterizations of the optical potential  \citep{cooper,KD} and of the MFA  prescriptions \citep{paloma,CK}.

The ratio BPT $\mathcal{R}_{\sigma}= \sigma_\mathrm{th}(\mathrm{QMC})/\sigma_\mathrm{th}(\mathrm{MFA})$, that quantifies the importance of nuclear correlations in the description of the QFS reactions, is represented in the middle panel of Fig.~\ref{fig:theoratios} and ranges from 0.6 to 1.
We have found that  the microscopic treatment of the overlaps has its biggest effect
on the evaluation of the theoretical cross section through the SFs.
We,  therefore, expect $\mathcal{R}_{\sigma}$ to be very close to $\mathcal{R}_\mathrm{\Sigma}$, which is confirmed when comparing the middle and upper panels.

The ratios $\mathcal{R}_{\sigma}$ exhibit the same moderate  dependence on $\mathrm{S}_\mathrm{N}$  as the $\mathcal{R}_\mathrm{\Sigma}$, the quenching  being
somewhat more significant for the knockout of the deficient species nucleon. For the symmetric case of $^{12}$C, the theoretical ratios still appear to 
indicate that protons are more correlated than neutrons.
Moroever,  the ratios $\mathcal{R}_{\sigma}$  are determined by the same delicate interplay between the radii of the parent and the residual nuclei
and the nucleon separation energy.
This is, in fact, a strong indication that the reaction mechanism does not probe directly the tail of the overlaps between the parent and residual nucleus. 
In addition, by the same physical arguments drawn for the SFs, we expect the quenching factors $\mathcal{R}_{S}$ to be smaller than unity,  independently of the interaction models.
The ratio $\mathcal{R}_{S}$  for (p,2p) from $^{12}\mathrm{C}$, also represented, is very close to $\mathcal{R}_\mathrm{\sigma}$ reflecting the ability of VMC to describe the data.

We represent in the lower panel of Fig.~\ref{fig:theoratios}  the ratios $\mathcal{R}_\mathrm{\sigma}$ and  $\mathcal{R}_S$ for nucleon removal from 
 $^{7}\mathrm{Li}$, $^{9}\mathrm{Li}$ and $^{10}\mathrm{Be}$, at 80-120 Mev/u, and 
 $^{12}\mathrm{C}$ measured at  250,1050, 2100 Mev/u.  The sp cross sections were taken from Refs.~\citep{Tostevin-02,grinyer12}
and  weighted by the QMC and CK SFs. 
The $\mathcal{R}_{S}$  are consistently smaller than  $\mathcal{R}_{\sigma}$.
For   $^{12}$C,  the deviation from  $\mathcal{R}_{\sigma}$ 
(which coincide for the 3 energies) appear to be dependant of the energy of the projectile for p- removal. 
Also, $\mathcal{R}_{\sigma}$ and  $\mathcal{R}_{S}$  appear to be  more quenched for proton than neutron knockout, yet with the exceptions of $^{10}\mathrm{Be}$ for the
former and   $^{12}$C for the later.
We point out, however, that in transfer reactions and in nucleon knockout reactions where both projectile and target are composite nuclei, the reaction mechanisms are substantially different from those of (p,pN) reactions, which prevents the conclusion that the cross section factorizes into SFs and single particle cross sections \citep{Deltuva-19},
being the factorization used merely  by convenience  \citep{grinyer12}.
Consequently, the clean link between the $\mathcal{R}_\mathrm{\Sigma}$ and $\mathcal{R}_{\sigma}$  
that exists for (p,pN) knockout reactions, shown in our work  for the first time, is not expected to hold for the removal analysis of the MSU experimental data \citep{grinyer12} and for transfer reactions. Accordingly, the behaviour of the quenching factors with respect to a given physical quantity is not directly related to the behavior of the ratio of the sums of the SFs with respect to the same quantity, and may have a very intricate interpretation. This conclusion is supported by the comparison of the lower panel  and upper panels of Fig.~\ref{fig:theoratios}: the location of the points in both cases is very different.

\section{Conclusion}

In conclusion, in the present letter we analyze, for the first time, (p,pN) reactions for A $\leq$ 12 and $(N-Z) \leq 3$ using state-of-the art  
nuclear structure and  3-body scattering formalisms, namely  QMC Wave Functions and the F/AGS reaction  theory. 
New QMC WFs used to describe $^{12}$C  and $^{11}$B yield results consistent with  experimental data of point rms radius,  electron scattering data analysis 
at high momentum and  (p,2p)  total cross sections at 400 MeV/u.

We show the inadequacy of the MFA to describe (p,pN) reactions due to contributions in the parent nucleus wave function of many-body partitions beyond the 
$(\mathrm{A}\!-\!1)+\mathrm{N}$, the only present in the MFA. This leads  necessarily to an overestimation of the cross sections,  independently of the interaction models.
Further, nuclear correlations are a key ingredient and, therefore, structure many-body effects must be taken into account. We show that the ratio between the partial sums of QMC and MFA QFS cross sections is very close to the ratio of the partial sum of SFs. Hence, one expects the quenching ratios to be determined by the same delicate interplay between the radii of the parent and the residual nuclei and the nucleon separation energy, as the ratio of the SFs. Last, $\mathcal{R}_\mathrm{\Sigma}$ and $\mathcal{R}_\mathrm{\sigma}$
show a moderate S$_N$ dependence and are smaller for the knockout of the more correlated deficient species nucleon. In the case 
of  the symmetric $^{12}$C,  the theoretical ratios  still appear to indicate that protons are more correlated than neutrons.
 
A consistent experimental program of transfer and knockout (with light and heavier targets) with proton
and electron probes for A $\leq$12 nuclei will be very useful to get further insight on the inadequacy of the MFA picture, on the structure of light nuclei. 


\section*{Acknowledgments}

R.C., E.C., A.A.,  A.M.  A.D. and  R.B.W. are supported by Funda\c c\~ao para a Ci\^encia e Tecnologia of Portugal, Grant No. PTD/FIS-NUC/2240/2014.
A.D. is supported  by the Alexander von Humboldt Foundation,  Grant No. LTU-1185721-HFST-E.
R.B.W. is supported by the US Department of Energy, Office of Nuclear Physics,
contract No. DE-AC02-06CH11357 and the NUCLEI SciDAC program; computing time was provided by the Laboratory
Computing Resource Center at Argonne National Laboratory.
We thank B. Jonson, P. Diaz-Fernandez and A. Heinz for reading the
manuscript.



\section*{References}

\begin{thebibliography}{99}
\bibitem{wiringa14}R. B. Wiringa, \textit{et al.},
Phys. Rev. C 89, 024305 (2014).
\bibitem{Carlson15}J. Carlson \textit{et al.},  Rev. Mod. Phys. 87, 1067 (2015).
\bibitem{Dickoff05}C. Barbieri, W.H. Dickoff, Int. Journal Mod. Phys A24, 2060 (2009)
\bibitem{Navratil08}S. Quaglioni and P. Navratil, Phys. Rev. Lett. 101, 092501 (2008).
\bibitem{Natasha13} N.\ K. \ Timofeyuk, Phys Rev. C 88, 044315 (2013).
\bibitem{Barranco17} F. Barranco  \textit{et al.}, Phys. Rev. Lett. 119, 082501 (2017).
\bibitem{Jensen} O. Jensen \textit{et al.}, Phys. Rev. Lett. 107, 032501 (2011).
\bibitem{Phandharipande} V. R. Pandharipande, I. Sick, P.K.A. de Witt Huberts, Rev. Mod. Phys. 69, 981 (1997).
\bibitem{nature}The CLAS Collaboration, Nature, 560, 617 (2018).
\bibitem{lapikas93}L. L\'apikas, Nucl. Phys. A 553,297 (1993).
\bibitem{lapikas99}L. L\'apikas, J.Wesseling, R. B. Wiringa, Phys. Rev. Lett. 82, 4404 (2009). 
\bibitem{Rhoe}D. Rohe \textit{et al.}, 
Phys. Rev. Lett. 93, 182501 (2004).
\bibitem{Subedi}R. Subedi \textit{et al.}, 
Science  320, 1475 (2009).
\bibitem{Kramer01}G. J. Kramer, H. P. Block, L. Lapikas, Nucl. Phys A
679, 267 (2001).
\bibitem{Tostevin-02} B.A. Brown, \textit{et al.},
Phys. Rev. C 65, 061601(R) (2002).
\bibitem{Tostevin-14} J.A. Tostevin and A. Gade, Phys. Rev. C 90, 057602 (2014).
\bibitem{grinyer12}G. F. Grinyer \textit{et al.}, Phys. Rev. C 86, 024315 (2012).
\bibitem{leyla} L. Atar,  \textit{et al.}, Phys. Rev. Lett. 120, 052501 (2018).
\bibitem{paloma} P. D\'iaz Fern\'andez  \textit{et al.}, 
Phys. Rev. C 97, 024311 (2018).
\bibitem{Nollett11}K. M. Nollett and R. B. Wiringa, Phys. Rev.  C 83, 0410001(R) (2011). 
\bibitem{Lee} J. Lee \textit{et al.}, Phys. Rev. Lett. 104, 112701  (2010).
\bibitem{flavignyPRL} F. Flavigny  \textit{et al.}, Phys. Rev. Lett. 110, 122503 (2013).
\bibitem{brida11}I. Brida, Steven C. Pieper and R. B. Wiringa, Phys. Rev. C 84, 024319 (2011).
\bibitem{Moro} M. G\'omez-Ramos and A.M. Moro, Phys. Lett B, 785, 511 (2018).
\bibitem{Bortignon-16} P. F. Bortignon and R. A. Broglia, Eur. Phys J. A  52, 64 (2016).
\bibitem{panin}V. Panin, \textit{et al.}, Phys. Lett. B 753, 204 (2016).
\bibitem{theorFad}L. D. Faddeev. Zh. Eksp. Teor. Fiz. 39, 1459 (1960).
\bibitem{alt:67a} E.~O. Alt, P. Grassberger, and W. Sandhas, Nucl.~Phys. {\bf B2},  167  (1967).
\bibitem{Deltuva-19}A. Deltuva, Phys Rev C 99, 024613 (2019).
\bibitem{Crespo-19}R. Crespo, E. Cravo, A. Deltuva, Phys Rev C 19, 054622 (2019).
\bibitem{cravo}E. Cravo, R. Crespo, A. Deltuva, Phys. Rev. C 93, 054612 (2016).
\bibitem{crespoJPG}R. Crespo \textit{et al.}, J. Phys: Conf Series, 966, 012056 (2018).
\bibitem{Witala06} H. Witala, J. Golak, R. Skibinski, Phys. Lett. B {\bf 634}, 374 (2006).
\bibitem{Witala11}H. Witala \textit{et al.}, Few-Body Systems 49, 1-4, pp 61-64 (2011).
\bibitem{KD} A. J. Koning and J. P. Delaroche, Nucl. Phys,
A713, 231 (2003).
\bibitem{cooper} E. D. Cooper \textit{et al.}, Phys. Rev. C 47, 297 (1993).
\bibitem{wiringa06}R. B. Wiringa, Phys. Rev. C 73, 034317 (2006).
\bibitem{piarulli18} M. Piarulli, \textit{et al.}, Phys. Rev. Lett. 120, 052503 (2018).
\bibitem{baroni18} M. Baroni, \textit{et al.}, Phys. Rev. C 98, 044003 (2018).
\bibitem{Vries87}H. De Vries, C. W. De Jager, C. De Vries, At. Data and Nucl. Data Tables {\bf 36}, 495 (1987).
\bibitem{Lu13}Z.-T. Lu \textit{et al.},  Rev. Mod. Phys. {\bf 85}, 1383 (2013).
\bibitem{CK}S. Cohen, D.  Kurath, Nucl. Phys A101, 1 (1967); D. Kurath (private communication).
\bibitem{cm_corr}A. E. L. Dieperink and T. de Forest, Jr. Phys. Rev. C 10, 543 (1974).
\bibitem{Steenhoven}G. van der Steenhoven, \textit{et al.}, Nucl. Phys.
A 480, 547, (1988). 
\end{thebibliography}
\end{document}